\documentclass{article}

\usepackage{arxiv}

\usepackage[utf8]{inputenc} 
\usepackage[T1]{fontenc}    
\usepackage{hyperref}       
\usepackage{url}            
\usepackage{booktabs}       
\usepackage{amsfonts}       
\usepackage{nicefrac}       
\usepackage{microtype}      
\usepackage{lipsum}
\usepackage{graphicx}
\graphicspath{ {./images/} }

\title{Predictive Analysis for Detection of Human Neck Postures using a robust integration of kinetics and kinematics}

\author{
 Korupalli V Rajesh Kumar\\
   Research Scholar \\
  School of Electronics Engineering\\
  Vellore Institute of Technology Chennai Campus\\
  Chennai, India 600127 \\
  \texttt{v.rajeshkumar2016@vitstudent.ac.in} \\
   \And
 Susan Elias \\
  Associate Professor \\
  School of Electronics Engineering\\
  Vellore Institute of Technology Chennai Campus\\
  Chennai, India 600127 \\
  \texttt{susan.elias@vit.ac.in} \\
}

\begin{document}
\maketitle
\begin{abstract}
Human neck postures and movements need to be monitored, measured, quantified and analyzed, as a preventive measure in healthcare applications. Improper neck postures are an increasing source of neck musculoskeletal disorders,   requiring therapy and rehabilitation. The motivation for the research presented in this paper was the need to develop a notification mechanism for improper neck usage. Kinematic data captured by sensors have limitations in accurately classifying the neck postures. Hence, we propose an integrated use of kinematic and kinetic data to efficiently classify neck postures. Using machine learning algorithms we obtained 100\% accuracy in the predictive analysis of this data. The research analysis and discussions show that the kinetic data of the Hyoid muscles can accurately detect the neck posture given the corresponding kinematic data captured by the neck-band. The proposed robust platform for the integration of kinematic and kinetic data has enabled the design of a smart neck-band for the prevention of neck musculoskeletal disorders.
\end{abstract}


\section{Introduction and Related Work}
\label{S1}

Enabling technology to monitor, measure and manage human movements has been an active area of research with a wide spectrum of applications ranging from medical diagnostics, rehabilitation, sports, fitness, behaviour analysis and gait based bio-metrics. The historic landscape of research publications in the field, has presented the use of technology that are  vision based (video cameras), sensor based (Inertial Measurement Units IMUs), Infra-red and RADAR based innovations to study human movement {\cite{ribeiro2005human, kellokumpu2008human, biswas2011gesture, nandy2016cloth}}. Quantitative analysis have traditionally been carried out using either kinetic or kinematic data for various applications. Kinematic data is obtained using IMUs and has been used in the study of Musculoskeletal Disorders (MSD) \cite {balakrishnan2016extensive, thanathornwong2014system} and in gait recognition. On the other hand, kinetic data provides details of the force of the component in motion and helps to analyse the activation of muscles associated with the joint in motion. Kinetic data are computed from the signals obtained from Kinesiological Electro Myography (KEMG) device and these are quantitatively analysed for understanding muscle force and fatigue \cite{kadaba1989repeatability,pizzolato2017comparison,ahamed2016age, deslandes2008isokinetics,saracc2017evaluation}.
The motivation for the research presented in this paper, was to identify a robust methodology to integrate the kinetic and kinematic features for predictive analysis of human postures and movements. IMU-based technology is the most popular for kinematic data capture as the IMU is a versatile Micro-Electro-Mechanical System (MEMs)  based device \cite{ seel2014imu}. IMU data have been integrated earlier with several proprietary and open source based simulation platforms for analysis and visualization of movement data \cite{shiroma2016daily,moncada2014activity,nguyen2017daily,mifsud2014portable, crema2017imu, georgi2015recognizing,  chavarriaga2013opportunity}. Here the integration of IMU with OpenSim is presented and it is currently a focus area in top research laboratories as well. OpenSim is an free open source simulation and modelling tool developed at Stanford University (\url{https://opensim.stanford.edu/}). The in-built features, modeling capabilities and the contributions by the OpenSim community makes it a scalable and reliable tool for analysing human movement. Besides presenting the step-wise procedure to integrate IMU data with OpenSim, this paper presents a novel methodology of combining kinetic and kinematic data for generating insightful analysis of human movements. OpenSim provides details of muscle activation and supports joint modelling of kinetic and kinematic parameters. In the following subsections the detailed methodology  is presented. The results and discussions that follow will highlight the significance and applications of the proposed robust integration platform for movement analysis.

 \section{Motivation and Proposed work}
 \label{s2}
The goal of the proposed research is to present a methodology to measure and accurately identify the postures of the human neck, for prevention and rehabilitation of musculoskeletal disorders of the cervical region. Improper neck posture due to sedentary lifestyle is a major cause of Cervical spine dysfunction in all age groups ranging from children to the elderly. The neck region can also be affected due to improper neck postures that are inherent in certain kinds of occupations. Other lifestyle related activities, including sleeping in the sitting position during travel can trigger musculoskeletal problems around the neck. The motivation for the research was to propose a novel methodology to prevent disorders of the neck through timely detection and notifications. A sensor based smart neck-band that can precisely detect neck postures was designed  to monitor improper use of the neck region and to generate alert messages as a preventive measure \cite{rajesh2019patent}. This neck-band can also be used to take measurements of the range of motion of the neck regions during therapy and  rehabilitation \cite{kumar2020smart}. Research works related to the use of sensors to track movements by obtaining kinematic data have been presented extensively in literature under the field called Actigraphy.  These approaches have also been widely used for tracking movement through commercially available wrist worn fitness monitors. But in order to identify the postures of the neck, there are several challenges and limitation in using only kinematic data obtained using sensors. Hence we explored the possibility of integrating  kinetic and kinematic data for better accuracy in detection of neck postures. In this paper we present a robust integrated platform for predictive analysis of human neck postures using kinetic and kinematic data. 
 
\section{Materials and Methods}
\label{S3}
In this research, an IMU based device is used for acquiring the kinematic data of the neck, and the OpenSim simulation modeling tool is used to  generate the  kinetic data of the corresponding neck movements. Predictive Analysis to detect neck postures from the kinetic and kinematic data, is done using Machine Learning methods. In the following sections, the methods and materials used in this research are presented.

\subsection{Kinematic data acquisition using Smart Neck-Band }

\subsubsection{IMU Based Neck-Band:}
Inertial Measurement Unit (IMU) embedded in an elastic neck-band captures the kinematic data required for the analysis. IMU devices are available in  miniature-sizes, and can be used to design  wearable products. For this research,  Metawear CPRO, IMU device developed at MBIENTLAB  has been used. It has on-chip memory, processing unit, accelerometer, gyroscope sensor, magnetometer sensor, pressure sensor and temperature sensor. In addition to these sensors, this device has inbuilt Bluetooth support, to establish  communication  with the associated mobile application or any Bluetooth device to transmit the device data. This IMU device can stream data for 6 hours continuously using a 3.3 V coin sized battery and  is attached to a  wearable band with an adjustable strap to fit it firmly around the neck. This neck-band when integrated with the proposed predictive analysis can be referred to as a smart neck-band,  for its context-aware functionalities.\\

\begin{figure}[h]
\begin{center}
\centering

\includegraphics[width = 12cm, height = 7cm]{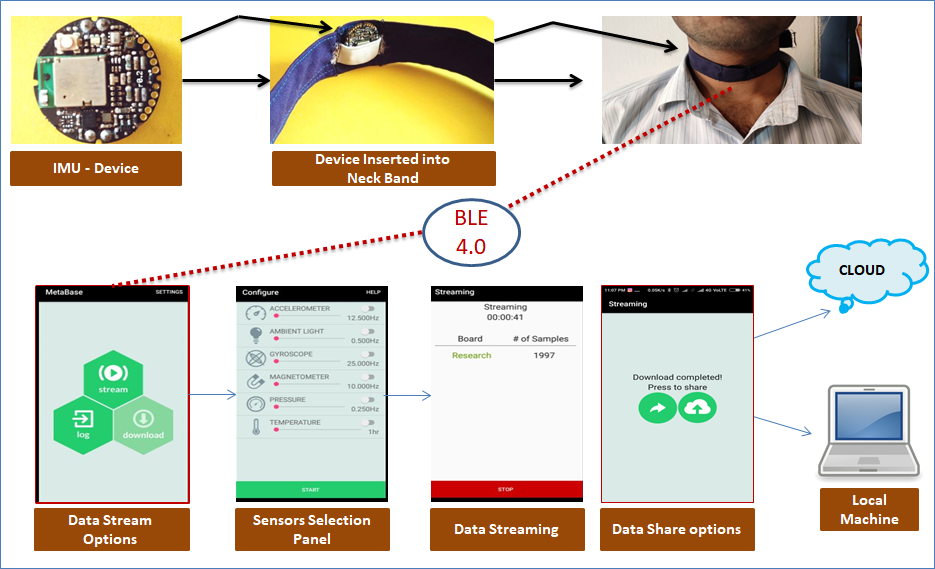}
\caption{Kinematic Data Acquisition Process }
\label{fig1}
\end{center}
\end{figure}

\subsubsection{Mobile Application- MetaBase:} 
The commercially available IMU device used in this research comes with an associated mobile application called MetaBase, developed by MBIENTLAB.  It is used to retrieve the recorded sensor data from the IMU device. MetaBase establishes a Bluetooth communication channel with the IMU hardware and receives the sensor data from the device. Once data transmission from the sensor to the mobile application is completed, the data can be shared to a cloud storage or to any local machine. Figure \ref{fig1} summarises the data acquisition process of the neck kinematics.

\subsubsection{Sensor Data Format:}
The IMU device, has in-built Accelerometer, Gyroscope and Magnetometer sensors to record various kinematic aspects of the movements.The format of the data captured by these sensors are given below:
\begin{itemize}
    \item Accelerometer sensor : epoch(ms),  time, elapsed(s), x(g), y(g), z(g)
    \item Gyroscope sensor  : epoch(ms),  time, elapsed(s), x(deg/s), y(deg/s), z(deg/s) 
    \item Magnetometer : epoch(ms),  time, elapsed(s), x(T), y(T), z(T) etc.
\end{itemize}
For the kinematic analysis of neck postures, dealt with in this research, the accelerometer sensor data was sufficient. The  accelerometer sensor was operated at 100Hz frequency with 8+/-g. \\

\subsection{Kinetic data generation using Neck Musculoskeletal Model }
Electro Myography (EMG) and Surface EMG (SEMG) are the standard  methods for muscle related data acquisition for movement analysis. The kinetic data obtained from EMG based methods are accurate, but are limited to laboratory based study. In order to carry out movement analysis  using both kinematic and kinetic data, a novel methodology  has been proposed in this paper. Here we focus on the detection of postures of the neck using an innovative approach. Instead of capturing kinetic data related to muscle activation using any of the EMG based methods, in the proposed approach kinetic data of the corresponding neck postures is generated using a Neck Musculoskeletal Simulation Model. The smart neck-band captures the real-world kinematic data of the neck postures and this data is used to generate the corresponding kinetic data relating to the muscles around the neck region using the OpenSim simulation tool. 

\subsubsection{OpenSim - Simulation Modeling Tool:}
\begin{figure}[h!]
\centering

\includegraphics[width = 12cm, height = 7cm]{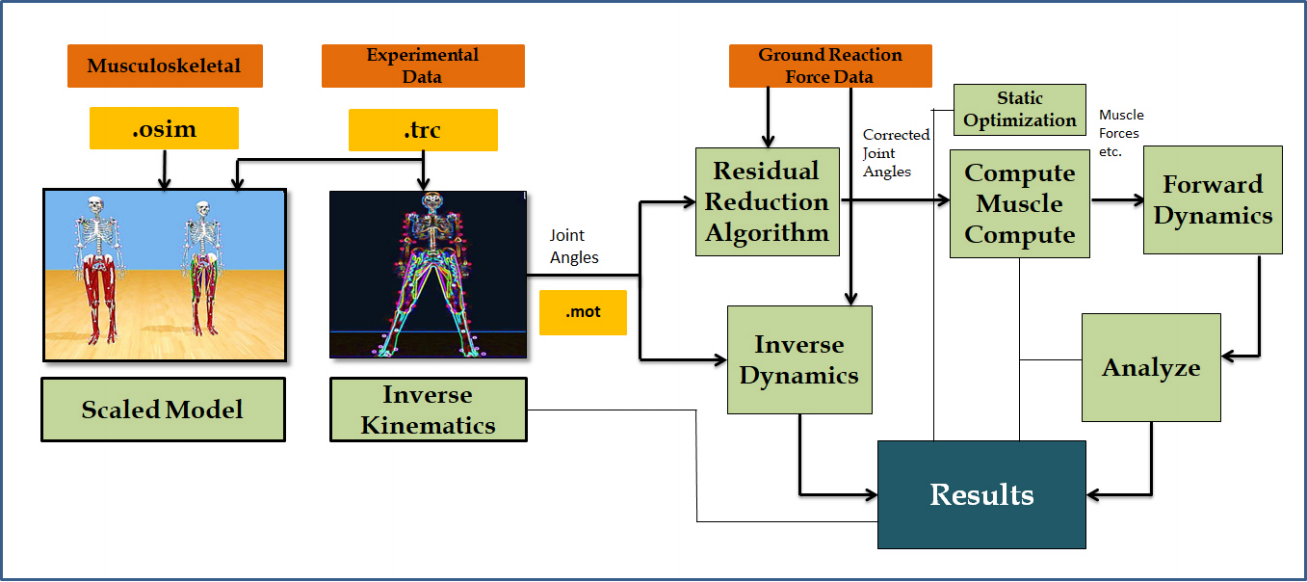}
\caption{Overview of OpenSim Functionalities}
\label{fig6}
\end{figure}

OpenSim is the most popular open-source tool, used to create and study human musculoskeletal models and  provides extensive data on kinematics and kinetics of human movement. In-built functionalities  such as Scale, Inverse Kinematics(IK), Inverse Dynamics (ID), Residual Reduction (RR), Static Optimization (STO), Computed Muscle Control (CMC) and Analyze,  provide support  to extract all information related to the muscles, tendons, joints kinetics and kinematics. An overview of the functionalities of OpenSim and the inbuilt tools is shown in Figure \ref{fig6} and detailed information  is made available by the OpenSim contributors \cite {delp2007opensim, seth2011opensim, thelen2006using, seth2018opensim}.\\ 

\subsubsection{Neck Musculoskeletal Model:} 
\begin{figure}[h]
\centering

\includegraphics[width = 12cm, height = 6cm]{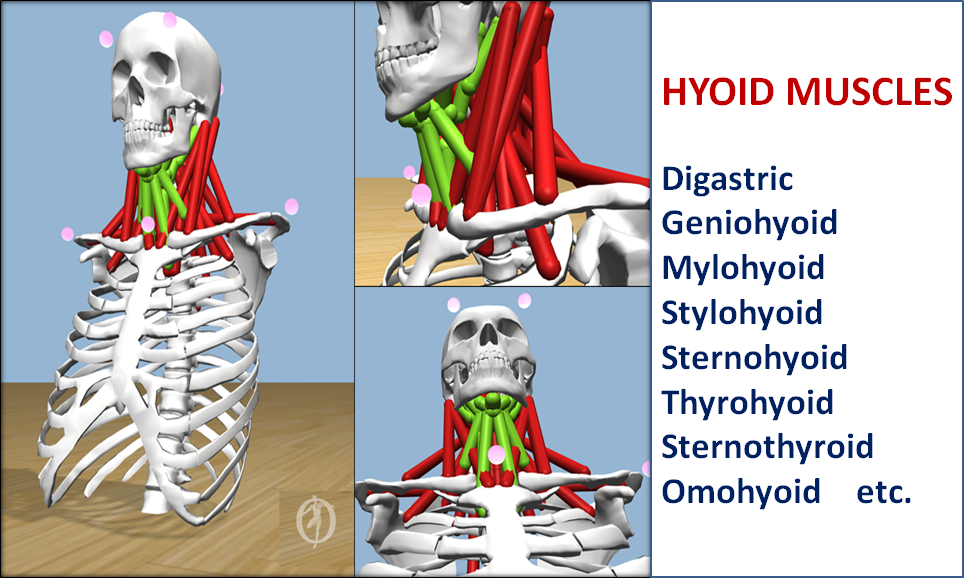}
\caption{Hyoid Muscles (in green colour)}
\label{fig12}
\end{figure}

In this research, the neck musculoskeletal model presented in \cite{mortensen2018inclusion} (Figure \ref{fig12}) was used. This is a fully flexible model for head and neck movements, and versatile compared to other models \cite{cazzola2017cervical}. The neck region consists of Cervical joints (C1-C7), sixty four muscles and various associated tendons and ligaments. The Hyoid muscles plays a vital role in supporting the neck movements and hence has a key role in the proposed predictive analysis. The Neck Musculoskeletal Model used in this research has integrated the Hyoid muscles and this was a big advantage for our experimental analysis. The kinetic data provided by OpenSim includes forces, length, power and activation levels of joints, muscles and tendons. With the neck  musculoskeletal model and research insights provided in \cite{mortensen2018inclusion, mortensen2018improved} used in simulation modeling. We used the kinetic data of the Hyoid muscles, extracted from the CMC tool during simulation for our research analysis. The Hyoid muscles are highlighted in green colour and shown in Figure \ref{fig12}. \\

\section{Experiments and Research Analysis }
\label{S4}
 In general the human neck has three-degrees of freedom, horizontal plane, vertical plane, and rolling of the head. All other asymmetric movements of the neck are variations and combinations of these three fundamental movements. At any point of time the neck will  be in one of the following nine static positions called neck postures or  it can move in any random order between these nine postures. The nine static positions or neck postures are:
\begin{itemize}
    \item  	Neck at Extreme Up (NU)
    \item  	Neck at Extreme Down (ND)
    \item  	Neck at Extreme Right (NR)
    \item  	Neck at Extreme Left (NL)
    \item  	Neck at Right Up (NRU)
    \item  	Neck at Right Down (NRD)
    \item  	Neck at Left Up (NLU)
    \item  	Neck at Left Down( NLD)
    \item  	Neck at Middle (NM).
\end{itemize}

The goal of the research presented in this paper is to design a methodology to detect neck postures by training the  Machine Learning algorithms using the kinematic and kinetic data.  In this research we used Random Forest, an ensemble learning method for the prediction and classification. The smart neck-band captures the real world data and integrates it with the OpenSim simulation platform. In order to effectively capture the neck kinematics, an experimental study was first carried out to determine the ideal location for the IMU device.  The IMU device was fixed onto the elastic band, and the neck-band could be worn in a manner that located the IMU device either in Front or at the Back of the neck region.  

\subsection{Experimental study for Location of IMU}
In this research, eight volunteers participated and details of their physical attributes are  given in Table \ref{tab1}.

\begin{table}[h]
\centering

\begin{tabular}{cccc}
\hline
\hline
\textbf{Gender} & \textbf{No.of Persons} & \textbf{Height} & \textbf{Weight} \\ \hline
Male            & 5                      & $\sim$5.3 cms   & +/- 75.6 kgs      \\ \hline
Female          & 3                      & $\sim$5.1 cms   & +/- 69.3 kgs      \\ \hline
\end{tabular}
\caption{Participants - Physical Attributes}
\label{tab1}
\end{table}

\begin{figure}[h]
\centering
\includegraphics[width = 10cm, height = 7cm]{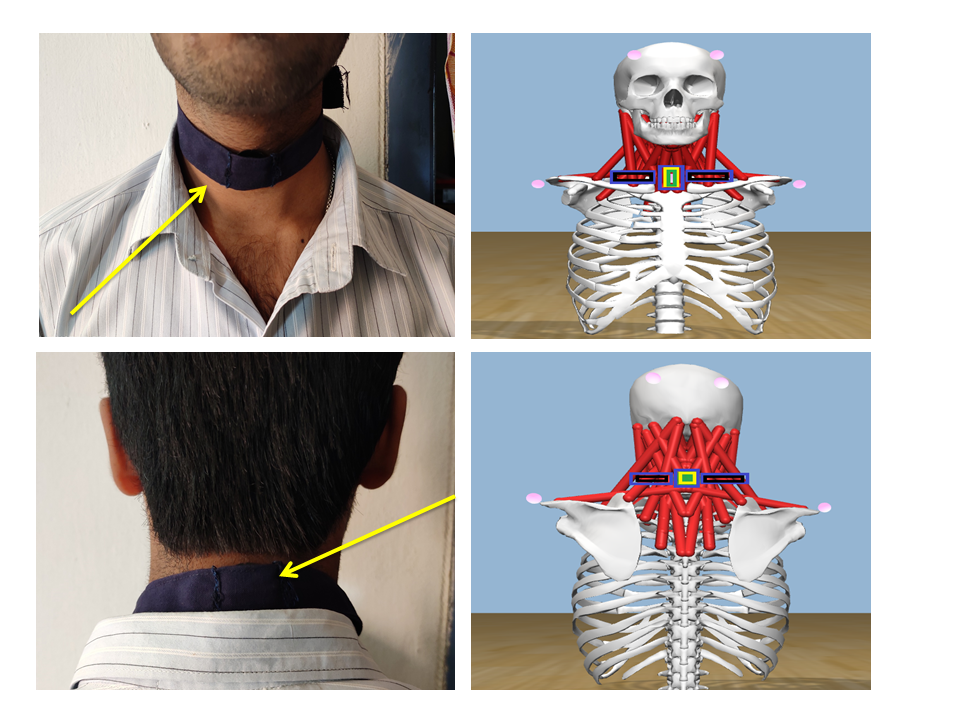}
\\
\caption{IMU located  FRONT \& BACK and corresponding Musculoskeletal Model}
\label{fig2}
\end{figure}

The subjects wore the neck-band and participated in  the research study. The participants were asked to keep their neck in the nine static positions for a duration ranging from 1 minute to 2 minutes based on their comfort level.  IMU data was recorded for all participants for each of the nine neck postures with the sensor located at FRONT side of the neck region and similarly for BACK side of neck. Figure \ref{fig2}, shows the kinematic and kinetic data extraction methods using IMU device and OpenSim simulation model for both Front and Back neck locations. Details of the dataset are presented in Table \ref{tab2}.

\begin{table}[h]
\centering
{%
\begin{tabular}{ccc}
\hline
\hline
\textbf{IMU - Device Placement} & \textbf{Total Time Duration} & \textbf{\begin{tabular}[c]{@{}c@{}}Dataset Size\\ (After Pre-processing)\end{tabular}} \\ \hline
Sensor Placed at Front Side & \begin{tabular}[c]{@{}c@{}}1080 sec\\ (Nine Static positions)\end{tabular} & \begin{tabular}[c]{@{}c@{}}1080 * 6\\ Rows  * Columns\end{tabular} \\ \hline
Sensor Placed at Back Side & \begin{tabular}[c]{@{}c@{}}1080 sec\\ (Nine Static positions)\end{tabular} & \begin{tabular}[c]{@{}c@{}}1080 * 6\\ Rows * Columns\end{tabular} \\ \hline
\end{tabular}%
}
\caption{Dataset Quantitative Parameters}
\label{tab2}
\end{table}

 Datasets were pre-processed and modularised based on the time stamp provided by the sensor and labelled manually. The observations in the dataset were also validated using the  video footage of the corresponding experimental study.  As part of the pro-processing task,  NaN’s(Not a Number) and  NA’s (Not Available) were interpreted, outliers were removed and data was normalised.

\subsection{Predictive Analysis using Random Forest Algorithm}
The data collected by keeping IMU device at the FRONT side of the neck was divided into training and testing data in the ratio of 75 : 25, and given as input to the Random Forest algorithm, to classify the  nine static neck positions and to  find the accuracy of the classification. Similarly the data collected by keeping the IMU device at the BACK side of the neck was similarly processed.

\begin{figure}[h]
\centering

\includegraphics[width = 12cm, height = 7cm]{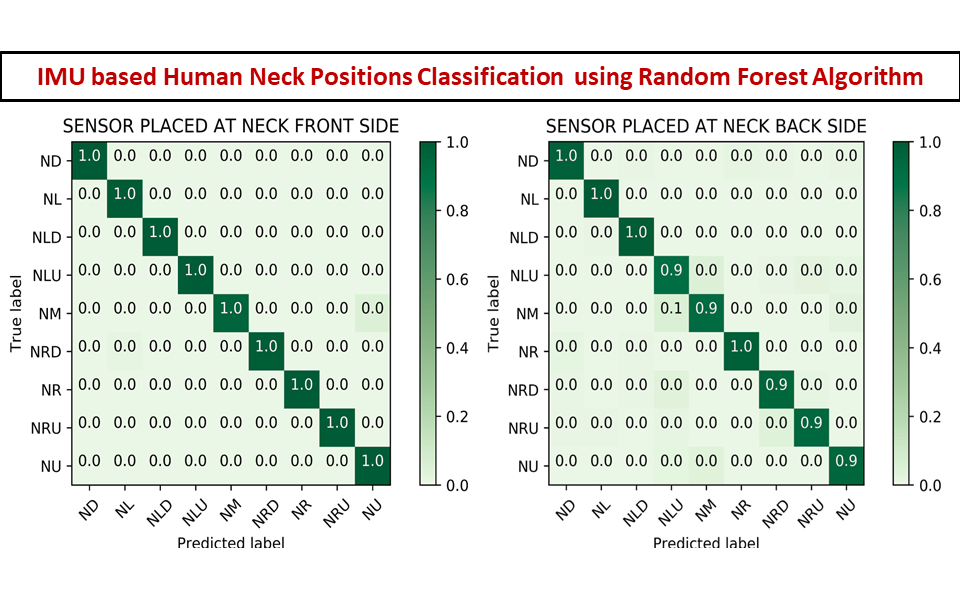}
\caption{Confusion Matrix -  Classification of Neck Positions}
\label{fig4}
\end{figure}

\begin{figure}[h]
\centering

\includegraphics[width = 10cm, height = 7cm]{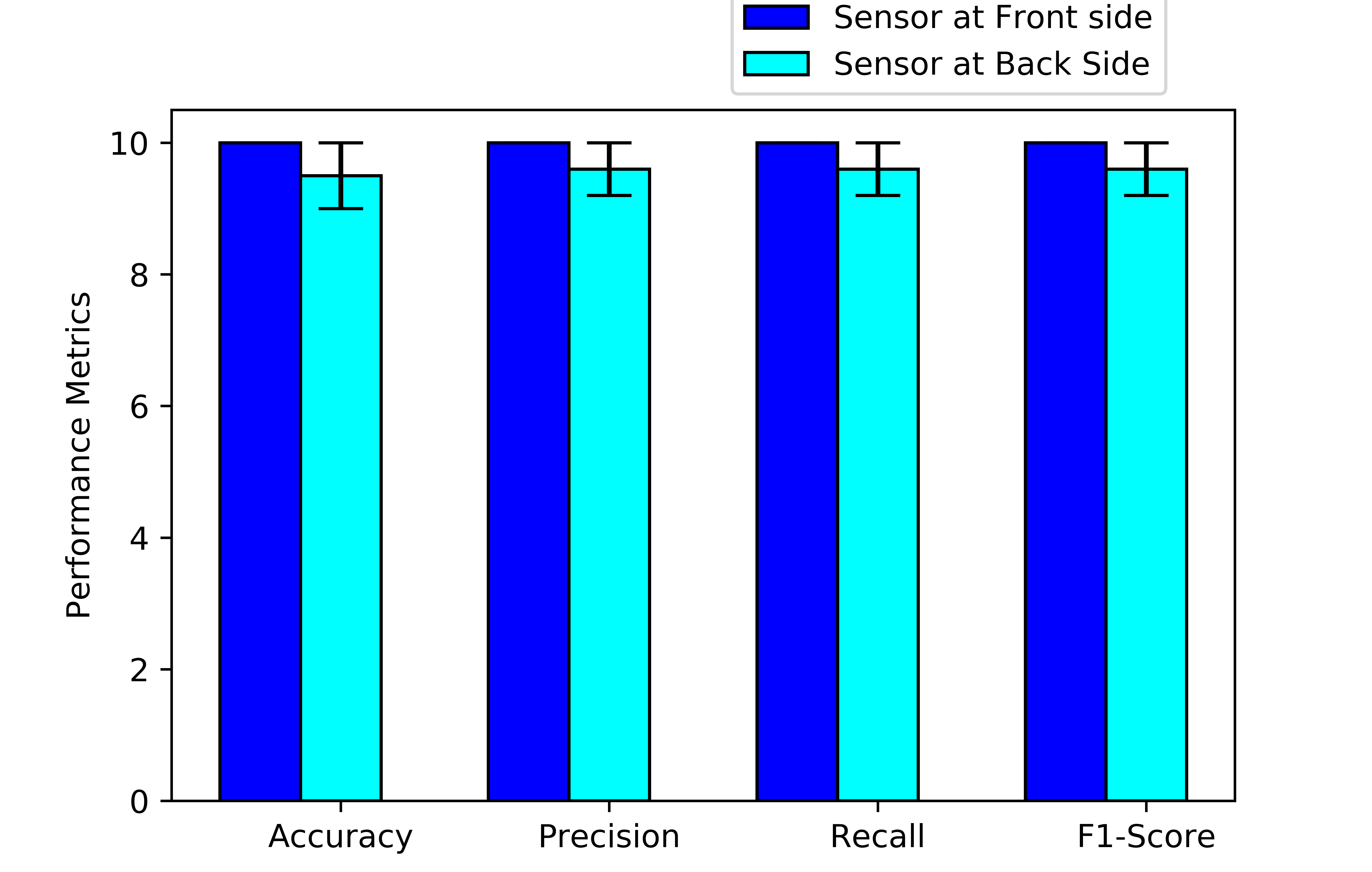}
\caption{Performance Metrics}
\label{fig5}
\end{figure}

The confusion matrix and performance metrics for the Front and Back location and  are shown in Figure \ref{fig4} and \ref{fig5}.  100\% accuracy was achieved for Front location based classification and 99\% for the Back location based classification. From this  research study we can infer that the ideal location for the IMU device during data capture is the Front side of the neck. This inference correlates with our decision to use the Hyoid muscles located in Front of the neck for accurate classification of neck postures. In this research paper, we proposed the idea of  generating kinetic data related to the Hyoid muscles and to use this data along with the associated kinematic data to detect neck posture accurately using classification techniques. This is presented in the following section. 

\section{Robust Integration of Kinematic and Kinetic Data}
\label{s4}
In this section we first present the procedure to integrate kinematic and kinetic data for the experiment research analysis. Integrating the IMU based kinematic data to the OpenSim simulation modeling platform is a  challenging and interesting task.

\subsection{IMU data integration with OpenSim}
\begin{figure}[h]
\centering

\includegraphics[width = 12cm, height = 6cm]{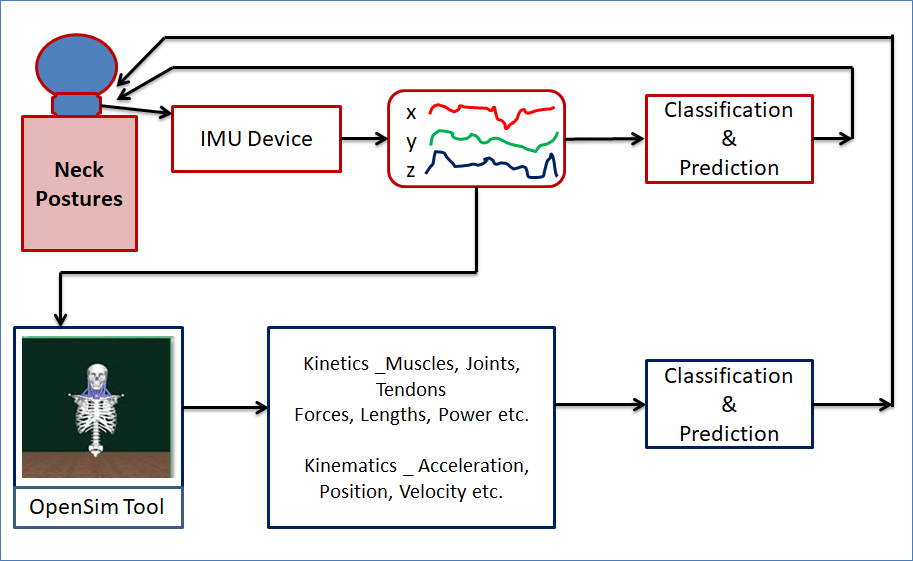}
\caption{IMU to OpenSim - Data Flow}
\label{fig7}
\end{figure}

\begin{itemize}
\item The IMU based accelerometer sensor data format provides three dimensional kinematic data (X,Y,Z).
\item To export IMU kinematic data into OpenSim simulation tool, mathematical and functional analysis is required. In OpenSim, a file with the extension .trc (Track Row Column) is used as an input file for the Inverse Kinematics (IK) tool, and using this information this tool provides joint movement data as a motion file with the extension .mot. 
\item The neck skeletal model has seven sets of markers around the skull and cervical region (four on the skull, one at Sternum Jugular Notch and two at acromioclavicular right and left joints ). In the front side of the neck, there is a marker called Sternum Jugular Notch (SJN). The IMU based kinematics data is mapped onto the data of SJN - X,Y,Z coordinates.  The other markers are calibrated according to the functional movements. The .trc file contains the details of these markers and it is the input file for the Inverse Kinematics (IK) and the  motion file (.mot) is obtained as the output. 
\item The information in this .mot file is fed as input to Computed Muscle Control (CMC) tool, which produces the data related to Neck kinematics, kinetics, joints, muscles, forces etc. \\
\item The functional integration of IMU data and OpenSim  is  shown in Figure \ref{fig7}.
\end{itemize}

\subsection{Simulation using Neck musculoskeletal model}

\begin{figure}[h]
\centering

\includegraphics[width = 14cm, height = 6cm]{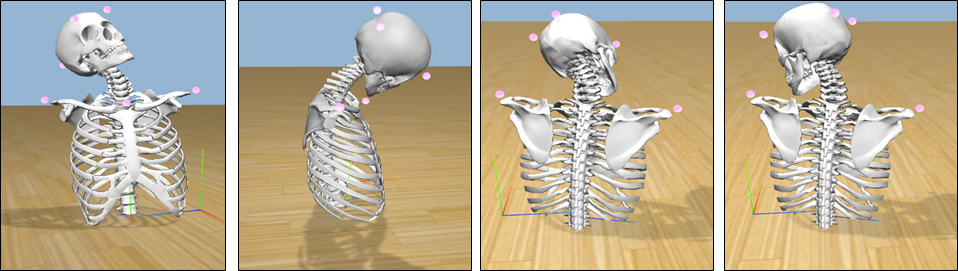}
\caption{Head and Neck Musculoskeletal Model}
\label{fig9}
\end{figure}

The Spine and Cervical joints C1, C2, C3, C4, C5, C6 and C7 play a pivotal role for any kind of neck movements. The Spine and Cervical variations in the OpenSim  for the various neck postures presented in Figure \ref{fig9}. 

\begin{figure}[h]
\centering

\includegraphics[width = 12cm, height = 6cm]{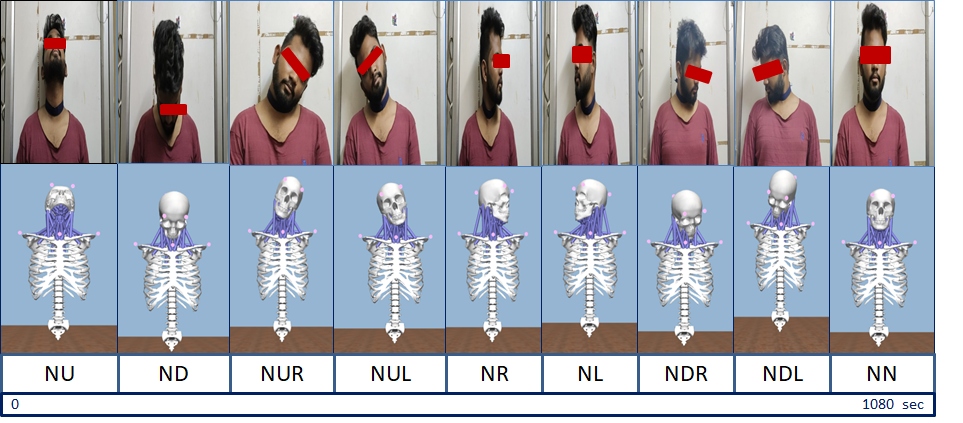}
\caption{ Subject Specific Neck Postures with OpenSim Musculoskeletal Model }
\label{fig8}
\end{figure}

In this research we recorded and analyzed kinematic data for all the subjects. In this section we present the simulation results and analysis for Subject ID: 04. The data collected from all the other subjects were used in training and testing of the  classification model. Figure \ref{fig8}, shows the  neck postures for research Subject ID: 4 and the corresponding  musculoskeletal model. The images were captured during the simulation of the Musculoskeletal neck model and this helped to correlate the dataset with the image of the neck posture.  

\subsection{Neck Musculoskeletal Model based Kinematic and Kinetic Data Analysis}

\begin{figure}[h]
\centering

\includegraphics[width = 12cm, height = 6cm]{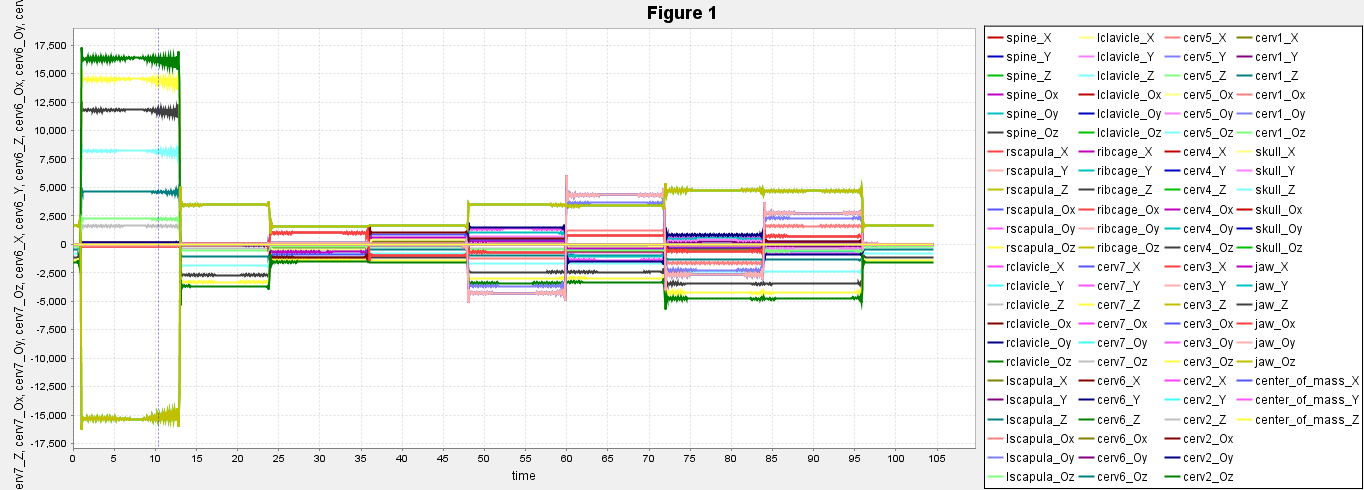}
\caption{Kinematics-Neck Acceleration }
\label{fig10}
\end{figure}

\begin{figure}[h]
\centering

\includegraphics[width = 12cm, height = 6cm]{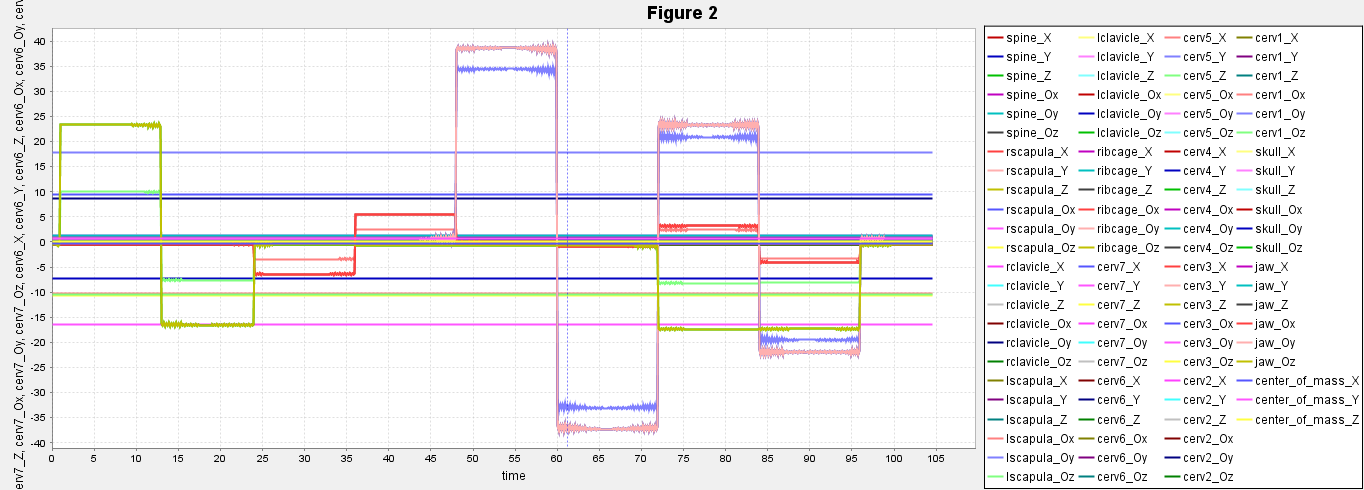}
\caption{Kinematics-Neck Position }
\label{fig11}
\end{figure}

OpenSim generates the Neck kinematics information based on the input data: Acceleration, Position and Velocity. We used  the Acceleration and Position data to classify and predict the human neck posture. From the Figures \ref{fig10} and \ref{fig11}, we can observe the response of the cervical joints and other associated joints during the experimentation task. The subjects changed the neck posture from one position to another after a time gap of about 120 sec, and with a  total of nine positions in the study, 1080 secs of data was captured. Figure \ref{fig10} shows the variations in the Neck Acceleration  with respective to joint kinematics and similarly Figure \ref{fig11} shows the variations in the Position of the neck. \\

\begin{figure}[h]
\centering

\includegraphics[width = 12cm, height = 6cm]{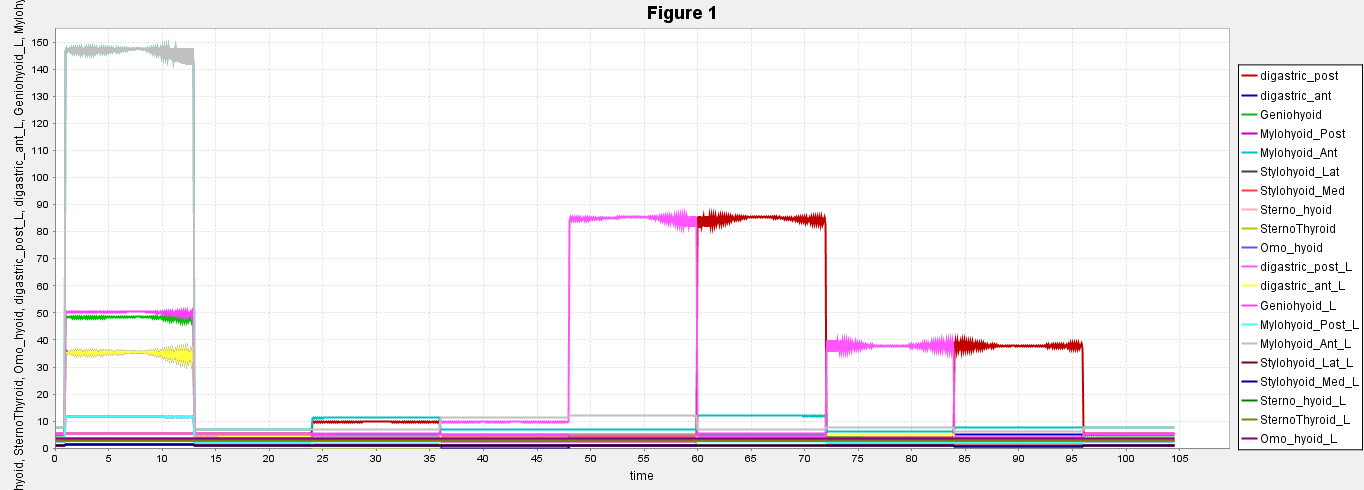}
\caption{Kinetics - Tendon Force }
\label{fig13}
\end{figure}

There are eight important sets of Hyoid muscles (shown in Figure \ref{fig12}) and many other associated hyoid muscles are attached with hyoid bone in neck region. These muscles help in providing free movement generation and flexibility to the neck \cite{mortensen2018inclusion}. The OpenSim based CMC tool provides kinetic information like  forces, activation, lengths etc. Using the CMC tool, kinetic data is extracted for the corresponding  kinematic data that was captured and integrated in OpenSim. Here, the response of the tendon forces of the Neck Hyoid  muscles were analysed  as shown in Figure \ref{fig13}.

\subsection{Predictive Analysis - Kinematics and Kinetics}

\begin{figure}[h]
\centering

\includegraphics[width = 10cm, height = 6cm]{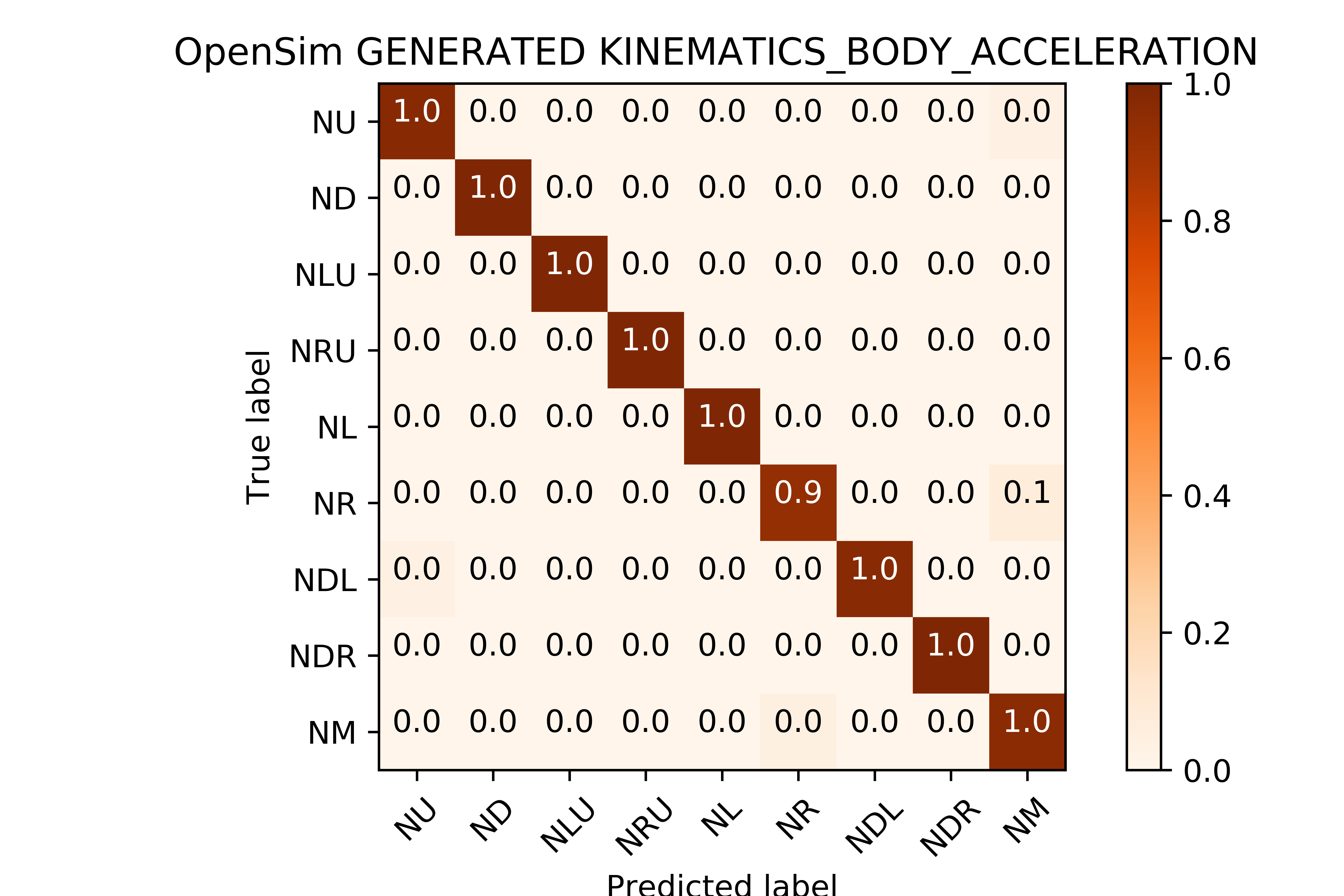}
\caption{Kinematics-Neck Acceleration }
\label{fig14}
\end{figure}

\begin{figure}[htb!]
\centering

\includegraphics[width = 10cm, height = 6cm]{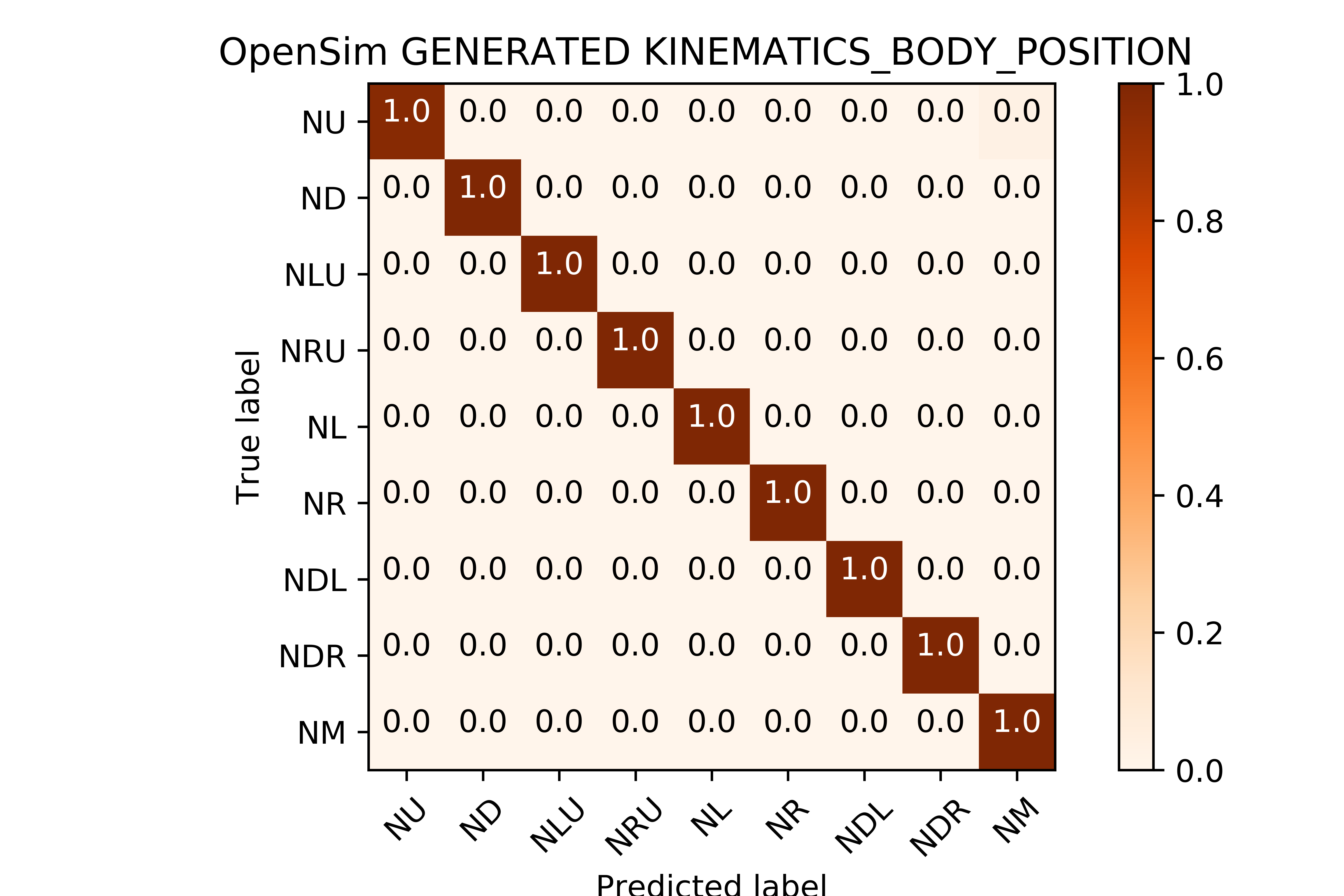}
\caption{Kinematics-Neck Position }
\label{fig15}
\end{figure}

The Random Forest Algorithm was used to classify and to predict the neck posture based on the kinematics and kinetics data generated by OpenSim. 
The Random Forest method achieved 100\% accuracy in the classification and prediction of neck postures using Neck acceleration and position data. These results are presented in Figures \ref{fig14} and \ref{fig15}. 

\begin{figure}[htb!]
\centering

\includegraphics[width = 10cm, height = 6cm]{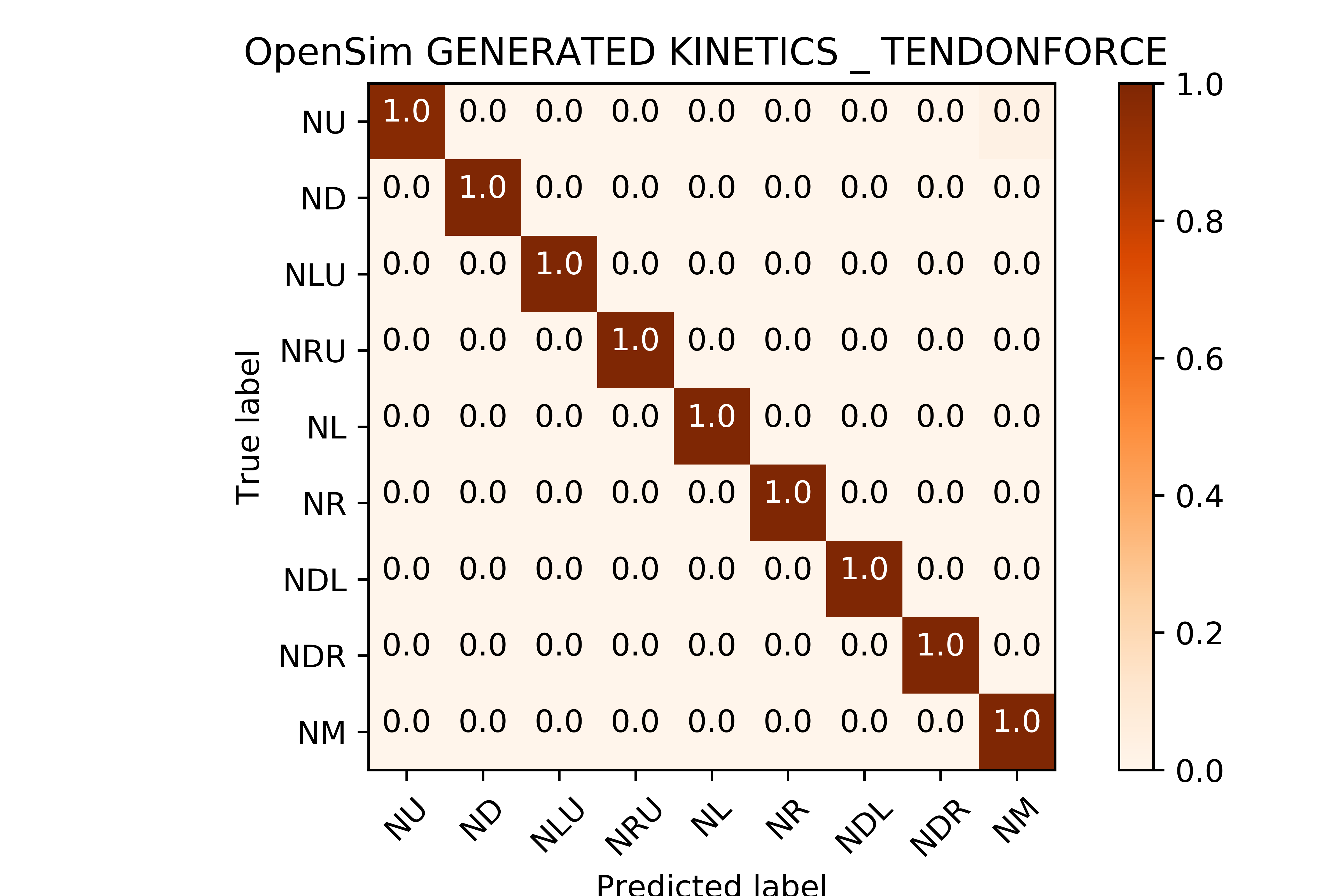}
\caption{Kinetics-Tendon Force}
\label{fig16}
\end{figure}

Similarly, Random Forest classifier predicted nine neck postures using the response of the tendon force data of Hyoid muscles and achieved 100\% accuracy in the prediction. The results of the kinetic tendon force classification is shown in Figure \ref{fig16}.

\section{Conclusion}
This paper presents a novel methodology to identify neck postures using kinetic and kinematic data. Improper neck postures can lead to neck cervical pain and musculoskeletal disorders. This research includes the design of a smart neck-band that captures kinematic data of the neck postures and movements. The OpenSim simulation tool and a neck musculoskeletal model, were used to simulate the related kinetic data for classification of the neck postures. The Machine Learning algorithms achieved 100\% accuracy in prediction of the neck postures and this innovative methodology can  support notification based applications for prevention of neck musculoskeletal disorders.

\bibliographystyle{hieeetr}
\bibliography{bibo}

\end{document}